\documentclass[a4paper,10pt,eqno]{article}
\usepackage{theorem}
\usepackage{latexsym,amssymb,amsfonts,amsmath}
\usepackage{cite}
\usepackage{color}
\newtheorem{thm}{Theorem}[section]
\newtheorem{lem}[thm]{Lemma}

\theoremheaderfont{\scshape}
\newcommand{\RM}{\mathbb{R}}

\newcommand{\EM}{\mathbb{E}}
\newcommand{\PM}{\mathbb{P}}

\newcommand{\qed}{\hfill $\Box$}

\newcommand{\ket}[1]{|#1\rangle}
\newcommand{\bra}[1]{\langle#1|}

\title{{\Large {\bf Localization of discrete time quantum walks on the glued trees}}}
\author{
{\small Yusuke Ide\footnote{To whom correspondence should be addressed. E-mail: ide@kanagawa-u.ac.jp}}\\
{\scriptsize Department of Information Systems Creation, 
Faculty of Engineering, 
Kanagawa University}\\
{\scriptsize Kanagawa, Yokohama 221-8686, Japan}\\
{\scriptsize e-mail: ide@kanagawa-u.ac.jp}\\
{\small Norio Konno}\\
{\scriptsize Department of Applied Mathematics, 
Faculty of Engineering, 
Yokohama National University}\\
{\scriptsize Hodogaya, Yokohama 240-8501, Japan}\\
{\scriptsize e-mail: konno@ynu.ac.jp}\\
{\small Etsuo Segawa}\\
{\scriptsize Graduate School of Information Science, 
Tohoku University}\\
{\scriptsize Aoba, Sendai 980-8579, Japan}\\
{\scriptsize e-mail: e-segawa@m.tohoku.ac.jp}\\
{\small Xin-Ping Xu}\\
{\scriptsize School of Physical Science and Technology, Soochow University}\\
{\scriptsize Suzhou 215006, China}\\
{\scriptsize Department of Physics and Astronomy, Seoul National University}\\
{\scriptsize Seoul 151-747, Korea}\\
{\scriptsize e-mail: xuxp@ihep.ac.cn}\\
}
\date{\empty }
\begin{document}
\maketitle

\par\noindent
\begin{small}
\par\noindent
{\bf Abstract}
\newline 
In this paper, we consider the time averaged distribution of discrete time quantum walks on the glued trees. In order to analyse the walks on the glued trees, we consider a reduction to the walks on path graphs. Using a spectral analysis of the Jacobi matrices defined by the corresponding random walks on the path graphs, we have spectral decomposition of the time evolution operator of the quantum walks. We find significant contributions of the eigenvalues $\pm 1$ of the Jacobi matrices to the time averaged limit distribution of the quantum walks. As a consequence we obtain lower bounds of the time averaged distribution. 
%\footnote[0]{
%{\it Abbr. title:} Localization of a DTQW on the path
%}
%\footnote[0]{
%{\it AMS 2000 subject classifications: }
%60F05, 60G50, 82B41, 81Q99
%}
%\footnote[0]{
%{\it PACS: } 
%03.67.Lx, 05.40.Fb, 02.50.Cw
%}
\end{small}

\setcounter{equation}{0}
%%%%%%%%%%%%%%%%%%%%%%%%%%%%%%%%%%%%%%%%%%

\section{Introduction}

The discrete time quantum walks (DTQWs) as quantum counterparts of the random walks which play important roles in various fields have been attractive research object in the last decade \cite{Kempe2003,Kendon2007,VAndraca2012,Konno2008b,AharonovEtAl2001,AmbainisEtAl2001,AhlbrechtEtAl2011, ManouchehriWang2013}. In the theory of quantum algorithm, quantum walks on various graphs also play important roles, for example, graph isomorphism testing and network characterization \cite{DouglasWang2008, BerryWang2010, BerryWang2011, RudingerEtAl2012}, search algorithms on the hypercube \cite{ShenviEtAl2003} or glued binary tree \cite{ChildsEtAl2003} and an algorithm for element distinctness on the Johnson graph \cite{Ambainis2004}. In these studies, the algorithms are often reduced to DTQWs on the path graphs. Therefore, investigations of DTQWs on the path graph corresponding to the original graphs are  important. Rohde {\em et al}.\ \cite{RohdeEtAl2011} studied periodic properties of entanglement for DTQW on the path determined by biased Hadamard coins numerically. Godsil \cite{Godsil2011} studied the time averaged distributions of continuous-time quantum walks on the path using the average mixing matrix. Ide {\em et al}.\ \cite{IdeKonnoSegawa2012} studied the time averaged distribution of DTQWs on the path graph which can be viewed as a quantization of random walks on the path. In this paper, we consider DTQWs on the path graphs corresponding to the random walks on the glued trees. We obtain lower bounds of the time averaged distribution of the DTQWs by using spectral analysis of the corresponding Jacobi matrices. 

The rest of this paper is organized as follows. The definition of our DTQW is given in Sect.\ 2 and main result of this paper is stated in Sect.\ 3. The remaining section (Sect.\ 4) is devoted to the proof of our result. 
%%%%%%%%%%%%%%%%%%%%%%%%%%%%%%%%%%%%%%%%%%

\section{Definition of the DTQW}

Let $T_{k}(n)$ be the $k$-ary tree on $\sum _{h=1}^{n}k^{h-1}$ vertices (height $=n$), i.e., the graph which is constructed inductively as follows: We start with a vertex called the ``root'' of $T_{k}(n)$ and set the height of the root $=1$. We add $k$ numbers of vertices and set the height of these vertices $=2$. Then we connect the root and all the vertices with its height $=2$. Similarly, for every vertex with its height $=h$, we add $k$ numbers of new vertices and set the height of these vertices $=h+1$. Then we connect the vertex with its height $=h$ and all the new $k$ vertices with their height $=h+1$. Note that there are $k^{h-1}$ numbers of vertices with their height $=h$. We repeat this procedure until all the vertices with its height $=n-1$ connect with $k$ numbers of new vertices with their height $=n$. We call each vertex with its height $=n$ ``leaf'' because the degree of these vertices equal one. Note that the degree of the root equals $k$, the degree of the leaves are one and the degree of all other vertices are $k+1$. 

In this paper, we consider the glued trees $G_{k}(2n)$ consisting of two $k$-ary trees $T_{k}^{1}(n)$ and $T_{k}^{2}(n)$. The glued tree $G_{k}(2n)$ is constructed as follows: Each leaf in $T_{k}^{1}(n)$ and $T_{k}^{2}(n)$ has $k$ numbers of ``potential edges''. We select a pair of potential edges $(e_{i},e_{j})$ at random where $e_{i}$ is a potential edge of a vertex $i$ in $T_{k}^{1}(n)$ and $e_{j}$ is that of $j$ in $T_{k}^{2}(n)$. After that we connect a pair of vertices $i$ and $j$ with an edge and erase the pair of potential edges $e_{i}$ and $e_{j}$. We continue this procedure until all the potential edges disappear. Note that the degree of the vertices in $G_{k}(2n)$ except for the two roots of $T_{k}^{1}(n)$ and $T_{k}^{2}(n)$ are $k+1$ and the degree of the roots are equal to $k$. 

In a quantum search algorithm \cite{ChildsEtAl2003}, it is known that the algorithms on the glued trees worked on the path graphs. For the Grover walks on the spidernets, it can be shown that there is a subspace which is isomorphic to a DTQW on the path graph. On this subspace, the Grover walk behaves as the DTQW which is called the Szegedy walk on the path graph (see \cite{KonnoObataSegawa2013} for more detail). Using similar argument, we can construct the Szegedy walks on the path graph corresponding to the Grover walks on the glued tree. Following these observations, we consider a reduction of $G_{k}(2n)$ on the path graph $P_{2n}$ on $2n$ numbers of vertices with the vertex set $V(P_{2n})=\{1,2,\cdots ,2n\}$ and the edge set $E(P_{2n})=\{(i,i+1) : i=1,2,\ldots 2n-1\}$. The results shown in this paper are restricted to the Szegedy walks on the path graph. But the results describe the behaviors of the corresponding Grover walks on the glued tree. Therefore it is useful to consider the Szegedy walks on the path graph.

First of all, we identify all the vertices with their height $=h$ in $T_{k}^{1}(n)$ and as the vertex $h$ in $P_{2n}$ and all the vertices with their height $=h$ in $T_{k}^{2}(n)$ as the vertex $2n-h+1$ in $P_{2n}$. Fig.\ 1  shows an example of the glued tree with $k=2$ and $n=3$ case. The figure also exhibits the corresponding path graph. 
%%%%%%%%%%%%%%%%%%%%%%%%%%%%%%%%%%%%%%%%%%%%%%%%%%%%%%%%%%%%%%%%%%%%%%%%%%%%%%%%%%
%%%%%%%%%%%%%%%%%%%%%%%%%%%%%%%%%%%%%%%%%%%%%%%%%%%%%%%%%%%%%%%%%%%%%%%%%%%%%%%%%%
\begin{figure}[htbp]
%\label{figtest}
\begin{center}
%WinTpicVersion4.26
\unitlength 0.1in
\begin{picture}( 34.0000, 19.0000)(  0.0000,-20.3500)
% LINE 2 0 3 0 Black White
% 40 3400 1000 2800 600 2800 1400 3400 1000 2800 600 2200 400 2800 600 2200 800 2200 1200 2800 1400 2800 1400 2200 1600 1600 1600 1000 1400 1000 1400 1600 1200 1600 800 1000 600 1000 600 1600 400 1000 600 400 1000 400 1000 1000 1400 1600 400 2200 1600 2200 800 1600 400 2200 400 1600 1200 1600 1200 2200 1200 1600 800 2200 1200 1600 800 2200 1600 1600 1600 2200 400 2200 800 1600 1600
% 
{\color[named]{Black}{%
\special{pn 8}%
\special{pa 3400 1000}%
\special{pa 2800 600}%
\special{fp}%
\special{pa 2800 1400}%
\special{pa 3400 1000}%
\special{fp}%
\special{pa 2800 600}%
\special{pa 2200 400}%
\special{fp}%
\special{pa 2800 600}%
\special{pa 2200 800}%
\special{fp}%
\special{pa 2200 1200}%
\special{pa 2800 1400}%
\special{fp}%
\special{pa 2800 1400}%
\special{pa 2200 1600}%
\special{fp}%
\special{pa 1600 1600}%
\special{pa 1000 1400}%
\special{fp}%
\special{pa 1000 1400}%
\special{pa 1600 1200}%
\special{fp}%
\special{pa 1600 800}%
\special{pa 1000 600}%
\special{fp}%
\special{pa 1000 600}%
\special{pa 1600 400}%
\special{fp}%
\special{pa 1000 600}%
\special{pa 400 1000}%
\special{fp}%
\special{pa 400 1000}%
\special{pa 1000 1400}%
\special{fp}%
\special{pa 1600 400}%
\special{pa 2200 1600}%
\special{fp}%
\special{pa 2200 800}%
\special{pa 1600 400}%
\special{fp}%
\special{pa 2200 400}%
\special{pa 1600 1200}%
\special{fp}%
\special{pa 1600 1200}%
\special{pa 2200 1200}%
\special{fp}%
\special{pa 1600 800}%
\special{pa 2200 1200}%
\special{fp}%
\special{pa 1600 800}%
\special{pa 2200 1600}%
\special{fp}%
\special{pa 1600 1600}%
\special{pa 2200 400}%
\special{fp}%
\special{pa 2200 800}%
\special{pa 1600 1600}%
\special{fp}%
}}%
% DOT 0 0 3 0 Black White
% 15 400 1000 1000 600 1000 1400 1600 1600 1600 1200 1600 800 1600 400 2200 400 2200 800 2200 1200 2200 1600 2800 1400 2800 600 3400 1000 3400 1000
% 
{\color[named]{Black}{%
\special{pn 4}%
\special{sh 1}%
\special{ar 400 1000 16 16 0  6.28318530717959E+0000}%
\special{sh 1}%
\special{ar 1000 600 16 16 0  6.28318530717959E+0000}%
\special{sh 1}%
\special{ar 1000 1400 16 16 0  6.28318530717959E+0000}%
\special{sh 1}%
\special{ar 1600 1600 16 16 0  6.28318530717959E+0000}%
\special{sh 1}%
\special{ar 1600 1200 16 16 0  6.28318530717959E+0000}%
\special{sh 1}%
\special{ar 1600 800 16 16 0  6.28318530717959E+0000}%
\special{sh 1}%
\special{ar 1600 400 16 16 0  6.28318530717959E+0000}%
\special{sh 1}%
\special{ar 2200 400 16 16 0  6.28318530717959E+0000}%
\special{sh 1}%
\special{ar 2200 800 16 16 0  6.28318530717959E+0000}%
\special{sh 1}%
\special{ar 2200 1200 16 16 0  6.28318530717959E+0000}%
\special{sh 1}%
\special{ar 2200 1600 16 16 0  6.28318530717959E+0000}%
\special{sh 1}%
\special{ar 2800 1400 16 16 0  6.28318530717959E+0000}%
\special{sh 1}%
\special{ar 2800 600 16 16 0  6.28318530717959E+0000}%
\special{sh 1}%
\special{ar 3400 1000 16 16 0  6.28318530717959E+0000}%
\special{sh 1}%
\special{ar 3400 1000 16 16 0  6.28318530717959E+0000}%
}}%
% DOT 0 0 3 0 Black White
% 7 3400 2000 2800 2000 2200 2000 1600 2000 1000 2000 400 2000 400 2000
% 
{\color[named]{Black}{%
\special{pn 4}%
\special{sh 1}%
\special{ar 3400 2000 16 16 0  6.28318530717959E+0000}%
\special{sh 1}%
\special{ar 2800 2000 16 16 0  6.28318530717959E+0000}%
\special{sh 1}%
\special{ar 2200 2000 16 16 0  6.28318530717959E+0000}%
\special{sh 1}%
\special{ar 1600 2000 16 16 0  6.28318530717959E+0000}%
\special{sh 1}%
\special{ar 1000 2000 16 16 0  6.28318530717959E+0000}%
\special{sh 1}%
\special{ar 400 2000 16 16 0  6.28318530717959E+0000}%
\special{sh 1}%
\special{ar 400 2000 16 16 0  6.28318530717959E+0000}%
}}%
% LINE 2 0 3 0 Black White
% 2 400 2000 3400 2000
% 
{\color[named]{Black}{%
\special{pn 8}%
\special{pa 400 2000}%
\special{pa 3400 2000}%
\special{fp}%
}}%
% LINE 2 2 3 0 Black White
% 12 400 2000 400 1000 1000 600 1000 2000 1600 2000 1600 400 2200 400 2200 2000 2800 2000 2800 600 3400 1000 3400 2000
% 
{\color[named]{Black}{%
\special{pn 8}%
\special{pa 400 2000}%
\special{pa 400 1000}%
\special{dt 0.045}%
\special{pa 1000 600}%
\special{pa 1000 2000}%
\special{dt 0.045}%
\special{pa 1600 2000}%
\special{pa 1600 400}%
\special{dt 0.045}%
\special{pa 2200 400}%
\special{pa 2200 2000}%
\special{dt 0.045}%
\special{pa 2800 2000}%
\special{pa 2800 600}%
\special{dt 0.045}%
\special{pa 3400 1000}%
\special{pa 3400 2000}%
\special{dt 0.045}%
}}%
% STR 2 0 3 0 Black White
% 4 500 550 500 600 2 0 0 0
% $T_{2}^{1}(3)$
\put(5.0000,-6.0000){\makebox(0,0)[lb]{$T_{2}^{1}(3)$}}%
% STR 2 0 3 0 Black White
% 4 3300 550 3300 600 3 0 0 0
% $T_{2}^{2}(3)$
\put(33.0000,-6.0000){\makebox(0,0)[rb]{$T_{2}^{2}(3)$}}%
% STR 2 0 3 0 Black White
% 4 400 2050 400 2100 5 0 0 0
% $1$
\put(4.0000,-21.0000){\makebox(0,0){$1$}}%
% STR 2 0 3 0 Black White
% 4 1000 2050 1000 2100 5 0 0 0
% $2$
\put(10.0000,-21.0000){\makebox(0,0){$2$}}%
% STR 2 0 3 0 Black White
% 4 1600 2050 1600 2100 5 0 0 0
% $3$
\put(16.0000,-21.0000){\makebox(0,0){$3$}}%
% STR 2 0 3 0 Black White
% 4 2200 2050 2200 2100 5 0 0 0
% $4$
\put(22.0000,-21.0000){\makebox(0,0){$4$}}%
% STR 2 0 3 0 Black White
% 4 2800 2050 2800 2100 5 0 0 0
% $5$
\put(28.0000,-21.0000){\makebox(0,0){$5$}}%
% STR 2 0 3 0 Black White
% 4 3400 2050 3400 2100 5 0 0 0
% $6$
\put(34.0000,-21.0000){\makebox(0,0){$6$}}%
% STR 2 0 3 0 Black White
% 4 200 1950 200 2000 5 0 0 0
% $P_{6}$
\put(2.0000,-20.0000){\makebox(0,0){$P_{6}$}}%
% STR 2 0 3 0 Black White
% 4 1900 150 1900 200 5 0 0 0
% $G_{2}(6)$
\put(19.0000,-2.0000){\makebox(0,0){$G_{2}(6)$}}%
\end{picture}%
\caption{A glued tree $G_{2}(6)$ consists of two $2$-ary trees with hight $=3$, $T_{2}^{1}(3)$ and $T_{2}^{2}(3)$. Each leaf of $T_{2}^{1}(3)$ is connected with  two randomly chosen leaves in $T_{2}^{2}(3)$. The glued tree $G_{2}(6)$ is reduced to the path graph $P_{6}$. }
\end{center}
\end{figure}
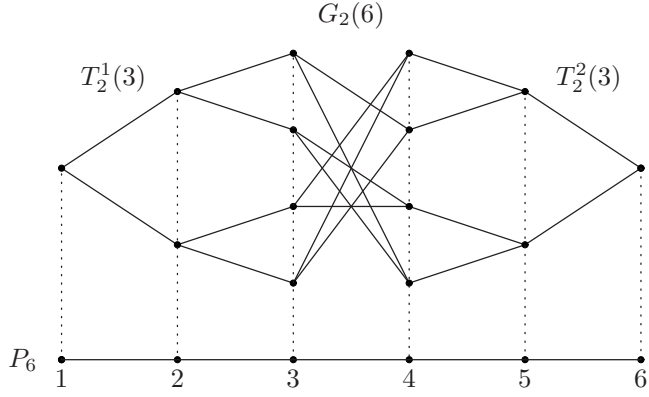
%%%%%%%%%%%%%%%%%%%%%%%%%%%%%%%%%%%%%%%%%%%%%%%%%%%%%%%%%%%%%%%%%%%%%%%%%%%%%%%%%%
%%%%%%%%%%%%%%%%%%%%%%%%%%%%%%%%%%%%%%%%%%%%%%%%%%%%%%%%%%%%%%%%%%%%%%%%%%%%%%%%%%

Recall that the simple random walk on $T_{k}^{1}(n)$ and $T_{k}^{2}(n)$ has the transition probabilities from a vertex $i$ to a neighboring  vertex $j$ as
\begin{align*}
p_{i,j}=
\begin{cases}
\frac{k}{k+1}, &\text{if the height of $j$ equals that of $i$ plus $1$},\\
\frac{1}{k+1}, &\text{if the height of $j$ equals that of $i$ minus $1$},
\end{cases}
\end{align*}
for the vertices $i$ except for the roots and 
\begin{align*}
p_{i,j}=\frac{1}{k},\quad \text{if the height of $j$ equals $2$},
\end{align*}
if the vertex $i$ is the root. Based on this fact, we consider the following random walk on $P_{2n}$ corresponding to $G_{k}(2n)$: Let $p=1/(k+1)$ and $q=k/(k+1)$. The transition probabilities of the random walk is
\begin{align*}
p_{i,i+1}&=
\begin{cases}
1, &\text{if $i=1$},\\
q, &\text{if $i=2,\ldots ,n$},\\
p, &\text{if $i=n+1,\ldots ,2n-1$},
\end{cases}\\
p_{i,i-1}&=
\begin{cases}
1, &\text{if $i=2n$},\\
q, &\text{if $i=n+1,\ldots ,2n-1$},\\
p, &\text{if $i=2,\ldots ,n$}.
\end{cases}
\end{align*}

Now we define corresponding DTQWs which we call the Szegedy walk \cite{IdeKonnoSegawa2012,Segawa2013,Szegedy2004} on $P_{2n}$ with general settings of transition probabilities $p$ and $q$ with $p+q=1$. This is the reduced DTQW on the path graph from the original Grover walk on the glued tree. Let $\mathcal{H}_{2n}= \mathrm{span}\{\ket{0,R}, \ket{1,L}, \ket{1,R},\ldots ,\ket{2n-1,L},\ket{2n-1,R},\ket{2n,L}\}$ be a Hilbert space with $\ket{i,J}=\ket{i}\otimes \ket{J}\ (i\in V(P_{2n}), J\in \{L,R\})$ the tensor product of elements of two orthonormal bases $\{\ket{i}:i\in V(P_{2n})\}$ for position of the walker and $\{\ket{L}={}^T [1,0], \ket{R}={}^T [0,1]\}$ for the chirality which means the direction of the motion of the walker where ${}^T \!\!A$ denotes the transpose of a matrix $A$.  Then we consider the time evolution operator $U^{(2n)}$ on $\mathcal{H}_{2n}$ defined by $U^{(2n)}=S^{(2n)}C^{(2n)}$ with the following coin operator $C^{(2n)}$ and shift operator $S^{(2n)}$:
\begin{align*}
C^{(2n)}
&=\ket{0}\bra{0}\otimes \ket{R}\bra{R}\\
&+\sum_{j=1}^{n}\ket{j}\bra{j}\otimes \left(2\ket{\phi_{1}}\bra{\phi_{1}}-I_{2}\right)
+\sum_{j=n+1}^{2n}\ket{j}\bra{j}\otimes \left(2\ket{\phi_{2}}\bra{\phi_{2}}-I_{2}\right)\\
&+\ket{2n}\bra{2n}\otimes \ket{L}\bra{L},\\
S^{(2n)}\ket{i,J}&=
\begin{cases}
\ket{i+1,L}&\text{if}\ \ J=R,\\ 
\ket{i-1,R}&\text{if}\ \ J=L,
\end{cases}
\end{align*}
where $\ket{\phi_{1}}=\sqrt{p}\ket{L}+\sqrt{q}\ket{R},\ \ket{\phi_{2}}=\sqrt{q}\ket{L}+\sqrt{p}\ket{R}$ and $I_{n}$ be the $n\times n$ identity matrix. 

Let $X_{t}^{(2n)}$ be the position of our quantum walker at time $t$. The probability that the walker with initial state $\ket{\psi}$ is found at time $t$ and the position $x$ is defined by 
\begin{eqnarray*}
\PM_{\ket{\psi}}(X_{t}^{(2n)}=x)=\left\lVert \left(\ket{x}\bra{x}\otimes I_{2}\right)\left(U^{(2n)}\right)^{t}\ket{\psi}\right\rVert^{2}.
\end{eqnarray*}
In this paper, we consider the DTQW starting from a vertex $i\in V(P_{2n})$ and choose the initial chirality state with equal probability, i.e., we choose the initial state as $\ket{\psi}_{1}=\ket{1}\otimes \ket{R}$ for $i=1$, $\ket{\psi}_{i}=\ket{i}\otimes \ket{L}$ or $\ket{\psi}_{i}=\ket{i}\otimes \ket{R}$ with probability $1/2$ for $2\leq i\leq 2n-1$, or $\ket{\psi}_{2n}=\ket{2n}\otimes \ket{L}$ for $i=2n$. For the sake of simplicity, we write $\PM_{i}(X_{t}^{(2n)}=x)$ for $\PM_{\ket{\psi}_{i}}(X_{t}^{(2n)}=x)$. Let the time averaged distribution 
\begin{eqnarray*}
\bar{p}_{i}^{(2n)}(x)=\lim _{T\to \infty }\EM\left[\frac{1}{T}\sum_{t=0}^{T-1}\PM_{i}(X_{t}^{(2n)}=x)\right],
\end{eqnarray*} 
where the expectation takes for the choice of the initial chirality state. For the sake of simplicity, we only consider $i,x\in \{1,\ldots, n\}$ case from now on.

%%%%%%%%%%%%%%%%%%%%%%%%%%%%%%%%%%%%%%%%%%

\section{Results}
In this section, we show our main result on the time averaged distribution $\bar{p}_{i}^{(2n)}(x)$. We allow $i,x\in \{1,\ldots, n\}$ to be fixed and diverging in $n$ cases. We have the following properties for $\bar{p}_{i}^{(2n)}(x)$: 
\begin{thm}\label{thmtimeavedist}
The time averaged distribution $\bar{p}_{i}^{(2n)}(x)$ is symmetric, i.e.,  $\bar{p}_{i}^{(2n)}(x)=\bar{p}_{2n+1-i}^{(2n)}(x)$ and $\bar{p}_{i}^{(2n)}(x)=\bar{p}_{i}^{(2n)}(2n+1-x)$ for $i,x\in \{1,\ldots, n\}$. The time averaged distribution has the following lower bounds:
\begin{align*}
\bar{p}_{i}^{(2n)}(x)\geq
\frac{(p-q)^{2}(q/p)^{i-1}}{2\left[\delta _{1}(i)+\left\{1-\delta _{1}(i)\right\}2q\right]\left\{2p-(q/p)^{n-1}\right\}^{2}}\left[\delta _{1}(x)+\{1-\delta _{1}(x)\}\frac{(q/p)^{x-1}}{q}\right].
\end{align*}
This leads to the following result:
\begin{enumerate}
\item
If $p>q$ then,
\begin{align*}
\liminf_{n\to \infty}\bar{p}_{i}^{(2n)}(x)\geq
\begin{cases}
\displaystyle\frac{(q/p)^{i+x-2}}{2\left[\delta _{1}(i)+\left\{1-\delta _{1}(i)\right\}2q\right]\left[\delta _{1}(x)+\left\{1-\delta _{1}(x)\right\}q\right]}\left(\frac{1-q/p}{2}\right)^{2}, &\text{if $i+x<\infty$},\\
0, &\text{otherwise}.
\end{cases}
\end{align*}
\item
If $p<q$ then,
\begin{align*}
\liminf_{n\to \infty}\bar{p}_{i}^{(2n)}(x)\geq
\begin{cases}
\displaystyle(p/q)^{k}\left(\frac{1-p/q}{2}\right)^{2}, &\text{if $\lim _{n\to \infty }\{2n-(i+x)\}=k\geq 0$},\\
0, &\text{otherwise}.
\end{cases}
\end{align*}
\end{enumerate}
\end{thm}

Note that $p=q=1/2$ case is included in the path graph case \cite{IdeKonnoSegawa2012}. Theorem \ref{thmtimeavedist} shows that if the random walker is likely to go to the two roots then the corresponding quantum walker localizes in vertices which are finitely close to the two roots. On the other hand, if the random walker is likely to go to the center of the glued tree then the quantum walker localizes in vertices which are finitely close to the leaves in $T^{1}(n)$ and $T^{2}(n)$. The proof of Theorem \ref{thmtimeavedist} is based on \cite{IdeKonnoSegawa2012,Szegedy2004}. In this proof, the eigenvalues and eigenvectors of the following $2n\times 2n$ finite Jacobi matrix $J_{2n}(p)$  induced by the random walk on $P_{2n}$ which $(i,j)$ component is determined by $\{J_{2n}(p)\}_{i,j}=\sqrt{p_{i,j}p_{j,i}}$, plays an important role: 
\begin{align*}
J_{2n}(p)=
\begin{bmatrix}
0 & \sqrt{p} & & & & & & & \\
\sqrt{p} & 0 & \sqrt{pq} & & & & & & \mbox{\smash{\huge\textit{O}}} \\
 & \sqrt{pq} & \ddots & \ddots & & & & & \\
 & & \ddots & \ddots & \sqrt{pq} & & & & & \\
 & & & \sqrt{pq} & 0 & q & & & \\
 & & & & q & 0 & \sqrt{pq} & & \\
 & & & & & \sqrt{pq} & \ddots & \ddots & \\
 & & & & & & \ddots & \ddots & \sqrt{pq} & \\
 & & & & & & & \sqrt{pq} & 0 & \sqrt{p} \\
\mbox{\smash{\huge\textit{O}}}  & & & & & & & & \sqrt{p} & 0
\end{bmatrix}.
\end{align*}
Indeed, as it is shown in Lemma \ref{lem:eigenU}, the eigenvalues and the eigenvectors of the time evolution operator $U^{(2n)}$ are given by that of $J_{2n}(p)$. On the other hand, the time averaged distribution is completely described 
by all the eigenvectors of $U^{(2n)}$. Unfortunately, as it is shown in Lemma \ref{lem:eigenJacobi}, we cannot obtain all of the eigenvalues of $J_{2n}(p)$ explicitly. The lower bounds in Theorem \ref{thmtimeavedist} are calculated by the eigenvectors corresponding to $\pm 1$ eigenvalues of $U^{(2n)}$. 
%%%%%%%%%%%%%%%%%%%%%%%%%%%%%%%%%%%%%%%%%%

\section{Proof of Theorem \ref{thmtimeavedist}}

In order to prove Theorem \ref{thmtimeavedist}, we have the following result for the eigenspace of the Jacobi matrices at first:
\begin{lem}[The determinantal formula for a symmetric Jacobi matrix]\label{lem:determinantal}
Let $J_{2n}$ be the following $2n\times 2n$ finite Jacobi matrix: 
\begin{align*}
J_{2n}=
\begin{bmatrix}
\alpha _{n} & \sqrt{w_{n-1}} & & & & & & & \\
\sqrt{w_{n-1}} & \ddots & \ddots & & & & & \mbox{\smash{\huge\textit{O}}} & \\
 & \ddots & \ddots & \sqrt{w_{1}} & & & & & \\
 & & \sqrt{w_{1}} & \alpha _{1} & \sqrt{w_{0}} & & & \\
 & & & \sqrt{w_{0}} & \alpha _{1} & \sqrt{w_{1}} & & \\
 & & & & \sqrt{w_{1}} & \ddots & \ddots & \\
 & & & & & \ddots & \ddots & \sqrt{w_{n-1}} \\
\mbox{\smash{\huge\textit{O}}}  & & & & & & \sqrt{w_{n-1}} & \alpha _{n}
\end{bmatrix}, 
\end{align*}
where $\alpha _{i}\in \RM$ for $i=1,\ldots n$ and $w_{i}\in (0,\infty )$ for $i=0,\ldots n-1$. Then the characteristic equation of $J_{2n}$, i.e., $\det(\lambda I_{2n}-J_{2n})=0$, is
\begin{align*}
\left\{(\lambda -\alpha _{1}-\sqrt{w_{0}})\det(E_{2})-w_{1}\det(E_{3})\right\}\left\{(\lambda -\alpha _{1}+\sqrt{w_{0}})\det(E_{2})-w_{1}\det(E_{3})\right\}=0,
\end{align*}
where $E_{k}$ be the following $(n-k+1)\times (n-k+1)$ matrix:
\begin{align*}
E_{k}=
\begin{bmatrix}
\lambda -\alpha _{n} & -\sqrt{w_{n-1}} & & & & \\
-\sqrt{w_{n-1}} & \ddots & \ddots & & \mbox{\smash{\huge\textit{O}}} & \\
 & \ddots & \ddots & \ddots & & \\
 & & \ddots & \ddots & -\sqrt{w_{k}} \\
\mbox{\smash{\huge\textit{O}}} & & & -\sqrt{w_{k}} & \lambda -\alpha _{k}
\end{bmatrix}.
\end{align*}
In addition, 
\newline
1. The eigenvector corresponding to the eigenvalue $\lambda $ with $(\lambda -\alpha _{1}-\sqrt{w_{0}})\det(E_{2})-w_{1}\det(E_{3})=0$ is 
\begin{align*}
{}^T \left[1, \frac{\det(E_{n})}{\sqrt{w_{n-1}}}, \frac{\det(E_{n-1})}{\sqrt{w_{n-2}w_{n-1}}},\ldots , \frac{\det(E_{2})}{\sqrt{w_{1}\cdots w_{n-1}}},  \frac{\det(E_{2})}{\sqrt{w_{1}\cdots w_{n-1}}}, \ldots , \frac{\det(E_{n-1})}{\sqrt{w_{n-2}w_{n-1}}}, \frac{\det(E_{n})}{\sqrt{w_{n-1}}}, 1\right].
\end{align*}
\newline
2. The eigenvector corresponding to the eigenvalue $\lambda $ with $(\lambda -\alpha _{1}+\sqrt{w_{0}})\det(E_{2})-w_{1}\det(E_{3})=0$ is 
\begin{align*}
{}^T \left[1, \frac{\det(E_{n})}{\sqrt{w_{n-1}}}, \frac{\det(E_{n-1})}{\sqrt{w_{n-2}w_{n-1}}},\ldots , \frac{\det(E_{2})}{\sqrt{w_{1}\cdots w_{n-1}}},  -\frac{\det(E_{2})}{\sqrt{w_{1}\cdots w_{n-1}}}, \ldots , -\frac{\det(E_{n-1})}{\sqrt{w_{n-2}w_{n-1}}}, -\frac{\det(E_{n})}{\sqrt{w_{n-1}}}, -1\right].
\end{align*}
\end{lem}
We should remark that every eigenvalue of $J_{2n}$ is simple (see e.g. \cite{HoraObata2007}). Therefore the eigenvectors are mutually orthogonal. 
\newline 
{\bf Proof of Lemma \ref{lem:determinantal}.}

Let 
\begin{align*}
\overline{E}_{k}=
\begin{bmatrix}
\lambda -\alpha _{k} & -\sqrt{w_{k}} & & & & \\
-\sqrt{w_{k}} & \ddots & \ddots & & \mbox{\smash{\huge\textit{O}}} & \\
 & \ddots & \ddots & \ddots & & \\
 & & \ddots & \ddots & -\sqrt{w_{n-1}} \\
\mbox{\smash{\huge\textit{O}}} & & & -\sqrt{w_{n-1}} & \lambda -\alpha _{n}
\end{bmatrix}.
\end{align*}
Note that by exchanging rows and colmuns, we have $\det(E_{k})=\det(\overline{E}_{k})$. 

By expanding $\det(\lambda I_{2n}-J_{2n})$ in the $n$-th row, we have
\begin{align*}
&\det(\lambda I_{2n}-J_{2n})\\
%&=
%\det
%\begin{bmatrix}
%\\
%\mbox{\smash{\huge $E_{3}$}} & -\sqrt{w_{2}} & & & &  \mbox{\smash{\huge\textit{O}}} & \\
%-\sqrt{w_{2}} & \lambda -\alpha _{2} & -\sqrt{w_{1}} & & & & \\
% & -\sqrt{w_{1}} & \lambda -\alpha _{1} & -\sqrt{w_{0}} & & & \\
% & & -\sqrt{w_{0}} & \lambda -\alpha _{1} & -\sqrt{w_{1}} & & \\
% & & & -\sqrt{w_{1}} & \lambda -\alpha _{2} &-\sqrt{w_{2}} & \\
% & & & & -\sqrt{w_{2}} & & \\
%\mbox{\smash{\huge\textit{O}}}  & & & & & \mbox{\smash{\huge $\overline{E}_{3}$}} 
%\end{bmatrix}\\
&=
\sqrt{w_{1}}\cdot 
\det
\begin{bmatrix}
\\
\mbox{\smash{\huge $E_{3}$}} & & & &  \mbox{\smash{\huge\textit{O}}} & \\
-\sqrt{w_{2}} & -\sqrt{w_{1}} & & & & \\
 & -\sqrt{w_{0}} & \lambda -\alpha _{1} & -\sqrt{w_{1}} & & \\
 & & -\sqrt{w_{1}} & \lambda -\alpha _{2} &-\sqrt{w_{2}} & \\
 & & & -\sqrt{w_{2}} & & \\
\mbox{\smash{\huge\textit{O}}}  & & & & \mbox{\smash{\huge $\overline{E}_{3}$}} 
\end{bmatrix}\\
&+(\lambda -\alpha _{1})\cdot
\det
\begin{bmatrix}
\\
\mbox{\smash{\huge $E_{2}$}} &  \mbox{\smash{\huge\textit{O}}} & \\
\\
\mbox{\smash{\huge\textit{O}}} & \mbox{\smash{\huge $\overline{E}_{1}$}} 
\end{bmatrix}
+\sqrt{w_{0}}\cdot 
\det
\begin{bmatrix}
\\
\mbox{\smash{\huge $E_{2}$}} & -\sqrt{w_{1}} & &  \mbox{\smash{\huge\textit{O}}} & \\
 & -\sqrt{w_{0}} & -\sqrt{w_{1}} & & \\
 & & \lambda -\alpha _{2} &-\sqrt{w_{2}} & \\
 & & -\sqrt{w_{2}} & & \\
\mbox{\smash{\huge\textit{O}}}  & & & \mbox{\smash{\huge $\overline{E}_{3}$}} 
\end{bmatrix}.
\end{align*}

On the other hand, repeating expansion of the determinants, we obtain
\begin{align*}
&\det
\begin{bmatrix}
\\
\mbox{\smash{\huge $E_{3}$}} & & & &  \mbox{\smash{\huge\textit{O}}} & \\
-\sqrt{w_{2}} & -\sqrt{w_{1}} & & & & \\
 & -\sqrt{w_{0}} & \lambda -\alpha _{1} & -\sqrt{w_{1}} & & \\
 & & -\sqrt{w_{1}} & \lambda -\alpha _{2} &-\sqrt{w_{2}} & \\
 & & & -\sqrt{w_{2}} & & \\
\mbox{\smash{\huge\textit{O}}}  & & & & \mbox{\smash{\huge $\overline{E}_{3}$}} 
\end{bmatrix}\\
%\sqrt{w_{0},\ldots ,w_{n-1}}\cdot
%\det
%\begin{bmatrix}
%\\
%\mbox{\smash{\huge $E_{3}$}} &  \mbox{\smash{\huge\textit{O}}} \\
%-\sqrt{w_{2}} & 0
%\end{bmatrix}
&=
-\sqrt{w_{1}}(\lambda -\alpha _{1})\cdot
\mathrm{det}
\begin{bmatrix}
\\
\mbox{\smash{\huge $E_{3}$}} &  \mbox{\smash{\huge\textit{O}}} & \\
\\
\mbox{\smash{\huge\textit{O}}} & \mbox{\smash{\huge $\overline{E}_{2}$}} 
\end{bmatrix}
+w_{1}\sqrt{w_{1}}\cdot
\mathrm{det}
\begin{bmatrix}
\\
\mbox{\smash{\huge $E_{3}$}} &  \mbox{\smash{\huge\textit{O}}} & \\
\\
\mbox{\smash{\huge\textit{O}}} & \mbox{\smash{\huge $\overline{E}_{3}$}} 
\end{bmatrix},
\end{align*}
and 
\begin{align*}
&\det
\begin{bmatrix}
\\
\mbox{\smash{\huge $E_{2}$}} & -\sqrt{w_{1}} & &  \mbox{\smash{\huge\textit{O}}} & \\
 & -\sqrt{w_{0}} & -\sqrt{w_{1}} & & \\
 & & \lambda -\alpha _{2} &-\sqrt{w_{2}} & \\
 & & -\sqrt{w_{2}} & & \\
\mbox{\smash{\huge\textit{O}}}  & & & \mbox{\smash{\huge $\overline{E}_{3}$}} 
\end{bmatrix}
%&=\sqrt{w_{0},\ldots ,w_{n-1}}\cdot
%\det
%\begin{bmatrix}
%0 & -\sqrt{w_{2}}\\
%\\
%\mbox{\smash{\huge\textit{O}}} & \mbox{\smash{\huge $\overline{E}_{3}$}}
%\end{bmatrix}
=
-\sqrt{w_{0}}\cdot 
\mathrm{det}
\begin{bmatrix}
\\
\mbox{\smash{\huge $E_{2}$}} &  \mbox{\smash{\huge\textit{O}}} & \\
\\
\mbox{\smash{\huge\textit{O}}} & \mbox{\smash{\huge $\overline{E}_{2}$}} 
\end{bmatrix}.
\end{align*}
Therefore, we have
\begin{align*}
&\det(\lambda I_{2n}-J_{2n})\\
&=-w_{1}(\lambda -\alpha _{1})\mathrm{det}(E_{2})\mathrm{det}(E_{3})+w_{1}^{2}\mathrm{det}(E_{3})^{2}+(\lambda -\alpha _{1})\det(E_{1})\det(E_{2})-w_{0}\mathrm{det}(E_{2})^{2}\\
&=(\lambda -\alpha _{1})^{2}\mathrm{det}(E_{2})^{2}-2w_{1}(\lambda -\alpha _{1})\mathrm{det}(E_{2})\mathrm{det}(E_{3})+w_{1}^{2}\mathrm{det}(E_{3})^{2}-w_{0}\mathrm{det}(E_{2})^{2}\\
&=\left\{(\lambda -\alpha _{1}-\sqrt{w_{0}})\det(E_{2})-w_{1}\det(E_{3})\right\}\left\{(\lambda -\alpha _{1}+\sqrt{w_{0}})\det(E_{2})-w_{1}\det(E_{3})\right\}.
\end{align*}
For the second equality, we use the following relation:
\begin{align*}
\mathrm{det}(E_{k})=(\lambda -\alpha _{k})\mathrm{det}(E_{k+1})-w_{k}\mathrm{det}(E_{k+2}),\ \text{for $k=2,\ldots ,n-1$}.
\end{align*}
This completes the first half of the proof.

We can easily check that 
\begin{align*}
\sqrt{w_{n-1}}\times \frac{\det(E_{n})}{\sqrt{w_{n-1}}}=\det(E_{n})=(\lambda -\alpha _{n})=(\lambda -\alpha _{n})\times 1,
\end{align*}
and
\begin{align*}
\sqrt{w_{k-1}}\times \frac{\det(E_{k})}{\sqrt{w_{k-1}\cdots w_{n-1}}}
&=\frac{1}{\sqrt{w_{k}\cdots w_{n-1}}}\times \left\{(\lambda -\alpha _{k})\mathrm{det}(E_{k+1})-w_{k}\mathrm{det}(E_{k+2})\right\}\\
&=(\lambda -\alpha _{k})\times \frac{\det(E_{k+1})}{\sqrt{w_{k}\cdots w_{n-1}}}-\sqrt{w_{k}}\times \frac{\det(E_{k+2})}{\sqrt{w_{k+1}\cdots w_{n-1}}},
\end{align*}
for $k=2,\ldots ,n-1$. 

From the condition $(\lambda -\alpha _{1}-\sqrt{w_{0}})\det(E_{2})-w_{1}\det(E_{3})=0$, we have
\begin{align*}
-\sqrt{w_{1}}\times \frac{\det(E_{3})}{\sqrt{w_{2}\cdots w_{n-1}}}+(\lambda -\alpha _{1})\times \frac{\det(E_{2})}{\sqrt{w_{1}\cdots w_{n-1}}}-\sqrt{w_{0}}\times \frac{\det(E_{2})}{\sqrt{w_{1}\cdots w_{n-1}}}=0.
\end{align*}
Similarly, from the condition $(\lambda -\alpha _{1}+\sqrt{w_{0}})\det(E_{2})-w_{1}\det(E_{3})=0$, we have
\begin{align*}
-\sqrt{w_{1}}\times \frac{\det(E_{3})}{\sqrt{w_{2}\cdots w_{n-1}}}+(\lambda -\alpha _{1})\times \frac{\det(E_{2})}{\sqrt{w_{1}\cdots w_{n-1}}}-\sqrt{w_{0}}\times \left(-\frac{\det(E_{2})}{\sqrt{w_{1}\cdots w_{n-1}}}\right)=0.
\end{align*}
 These conditions imply that the vectors described in the lemma are the corresponding eigenvectors. 
\qed

Now we apply Lemma \ref{lem:determinantal}  with parameters $\alpha _{1}=\cdots =\alpha _{n}=0$, $w_{0}=q^{2}$, $w_{2}=\cdots =w_{n-2}=pq$ and $w_{n-1}=p$ to the Jacobi matrix $J_{2n}(p)$. We obtain the following characteristic equation of $J_{2n}(p)$:
\begin{align*}
\left\{(\lambda -q)\det (E^{\prime }_{n-1})-pq\det (E^{\prime }_{n-2})\right\}\left\{(\lambda +q)\det (E^{\prime }_{n-1})-pq\det (E^{\prime }_{n-2})\right\}=0,
\end{align*}
where $E^{\prime }_{k}$ is the following $k\times k$ matrix:
\begin{align*}
E^{\prime }_{k}=
\begin{bmatrix}
\lambda  & -\sqrt{p} & & & & \\
-\sqrt{p} & \lambda & -\sqrt{pq} & & \mbox{\smash{\huge\textit{O}}} & \\
 & -\sqrt{pq} & \ddots & \ddots & & \\
 & & \ddots & \lambda & -\sqrt{pq} \\
\mbox{\smash{\huge\textit{O}}} & & & -\sqrt{pq} & \lambda 
\end{bmatrix}.
\end{align*}
Remark that for $k\geq 2$, 
\begin{align*}
\det (E^{\prime }_{k})&=\lambda \det (E^{\prime }_{k-1})-pq\det (E^{\prime }_{k-2})=\lambda \det (F_{k-1})-p\det (F_{k-2}),\\
\det (F_{k})&=\lambda \det (F_{k-1})-pq\det (F_{k-2}),
\end{align*}
where $F_{k}$ is the following $k\times k$ matrix:
\begin{align*}
F_{k}=
\begin{bmatrix}
\lambda  & -\sqrt{pq} & & & \\
-\sqrt{pq} & \ddots & \ddots & \mbox{\smash{\huge\textit{O}}} & \\
 & \ddots & \ddots & -\sqrt{pq} \\
\mbox{\smash{\huge\textit{O}}} & & -\sqrt{pq} & \lambda 
\end{bmatrix}.
\end{align*}
Using these facts, we can calculate
\begin{align*}
(\lambda \mp q)\det (E^{\prime }_{n-1})-pq\det (E^{\prime }_{n-2})=(\lambda \mp 1)\left\{\det (F_{n-1})\pm p\det (F_{n-2})\right\}.
\end{align*}
Therefore the eigenvalues of $J_{2n}(p)$ are $\lambda =\pm 1$ and $\lambda $ satisfying $\det (F_{n-1})\pm p\det (F_{n-2})=0$. 

For $\lambda =\pm 1$ case, we have
\begin{align*}
\det (E^{\prime }_{k})&=\pm \det (F_{k-1})-p\det (F_{k-2}),\\
\det (F_{k})&=\pm \det (F_{k-1})-pq\det (F_{k-2}),
\end{align*}
with $\det (F_{1})=\pm 1$ and $\det (F_{0})=1$. This implies
\begin{align*}
\det (E^{\prime }_{k})&=(\pm 1)^{k}q^{k-1},\ \text{for $k\geq 1$}.
\end{align*}

On the other hand, for $\lambda \neq \pm 1$ case, we identify $\det (F_{k})$ as $\sqrt{pq}^{k}\tilde{U}_{k}(\lambda /\sqrt{pq})$ where $\tilde{U}_{k}(x)$ is the (monic) Chebyshev polynomial of the second kind, i.e., the series of polynomials satisfying the following recurrence relation:
\begin{align*}
\tilde{U}_{0}(x)&=1,\\
\tilde{U}_{1}(x)&=x,\\
\tilde{U}_{k}(x)&=x\tilde{U}_{k-1}(x)-\tilde{U}_{k-2}(x),\ \text{for $k\geq 2$}.
\end{align*}
In this case, we obtain
\begin{align*}
\det (F_{n-1})\pm p\det (F_{n-2})&=\sqrt{p}\sqrt{pq}^{n-2}\left\{\sqrt{q}\tilde{U}_{n-1}(\lambda /\sqrt{pq})\pm \sqrt{p}\tilde{U}_{n-2}(\lambda /\sqrt{pq})\right\},\\
\det (E^{\prime }_{k})&=\frac{\sqrt{pq}^{k}}{q}\left\{q\tilde{U}_{k}(\lambda /\sqrt{pq})- p\tilde{U}_{k-2}(\lambda /\sqrt{pq})\right\}, \text{for $k\geq 1$},
\end{align*}
with $\tilde{U}_{-1}(x)=0$. 
Combining these results with Lemma \ref{lem:determinantal}, we have the following lemma for the eigen space of $J_{2n}(p)$:
\begin{lem}\label{lem:eigenJacobi}
Let $\lambda $ be the eigenvalue of $J_{2n}(p)$ and $\mathbf{v}_{\lambda }$ be the corresponding eigenvector. Then we have $\lambda=\pm 1$ and the remaining eigenvalues $\lambda $ satisfy $\sqrt{q}\tilde{U}_{n-1}(\lambda /\sqrt{pq})\}\pm \sqrt{p}\tilde{U}_{n-2}(\lambda /\sqrt{pq})=0$. The i-th component  $\mathbf{v}_{\lambda }(i)$ of the eigenvectors $\mathbf{v}_{\lambda }$ are the following:
\begin{enumerate}
\item
For $\lambda =\pm 1$,
\begin{align*}
\mathbf{v}_{\pm 1}(i)&=
\begin{cases}
1, &\text{if $i=1$},\\
(\pm \sqrt{q/p})^{i-1}/\sqrt{q}, &\text{if $i=2,\ldots,n$},\\
\pm (\pm \sqrt{q/p})^{2n-i}/\sqrt{q}, &\text{if $i=n+1,\ldots,2n-1$},\\
\pm 1, &\text{if $i=2n$}.\\
\end{cases}\\
||\mathbf{v}_{\pm 1}||^{2}
&=
\begin{cases}
2(2n-1), &\text{if $p=q=1/2$},\\
\frac{2}{p-q}\left\{2p-(q/p)^{n-1}\right\}, &\text{if $p\neq q$}.\\
\end{cases}
\end{align*}
\item
For $\lambda \neq \pm 1$,
\begin{align*}
\mathbf{v}_{\lambda }(i)&=
\begin{cases}
1, &\text{if $i=1$},\\
\left\{q\tilde{U}_{i}(\lambda /\sqrt{pq})-p\tilde{U}_{i-2}(\lambda /\sqrt{pq})\right\}/\sqrt{q}, &\text{if $i=2,\ldots,n$},\\
\pm \left\{q\tilde{U}_{2n-i}(\lambda /\sqrt{pq})-p\tilde{U}_{2n-i-2}(\lambda /\sqrt{pq})\right\}/\sqrt{q}, &\text{if $i=n+1,\ldots,2n-1$},\\
\pm 1, &\text{if $i=2n$}.\\
\end{cases}\\
||\mathbf{v}_{\lambda }||^{2}
&=
\frac{2(1-\lambda ^{2})S_{n-1}}{q},\ \text{with $S_{k}=\sum_{i=0}^{k-1}\tilde{U}_{i}(\lambda /\sqrt{pq})$}. 
\end{align*}
\end{enumerate}
\end{lem}

The last part of Lemma \ref{lem:eigenJacobi}, i.e., $||\mathbf{v}_{\lambda }||^{2}$, is calculated as follows:
\begin{align*}
||\mathbf{v}_{\lambda }||^{2}
&=2\left[1+q\tilde{U}_{1}^{2}(\lambda /\sqrt{pq})+\frac{1}{q}\sum_{i=2}^{n-1}\left\{q\tilde{U}_{i}(\lambda /\sqrt{pq})-p\tilde{U}_{i-2}(\lambda /\sqrt{pq})\right\}^{2}\right]\\
&=2\left\{1+q\tilde{U}_{1}^{2}(\lambda /\sqrt{pq})+\sum_{i=2}^{n-1}\tilde{U}_{i}^{2}(\lambda /\sqrt{pq})-\frac{\lambda ^{2}}{q}\sum_{i=2}^{n-2}\tilde{U}_{i}^{2}(\lambda /\sqrt{pq})+\frac{p}{q}\sum_{i=2}^{n-3}\tilde{U}_{i}^{2}(\lambda /\sqrt{pq})\right\}.
\end{align*}
Here we use $\lambda ^{2}\tilde{U}_{i-1}^{2}(\lambda /\sqrt{pq})=pq\tilde{U}_{i}^{2}(\lambda /\sqrt{pq})+pq\tilde{U}_{i-2}^{2}(\lambda /\sqrt{pq})+2pq\tilde{U}_{i}(\lambda /\sqrt{pq})\tilde{U}_{i-2}(\lambda /\sqrt{pq})$ obtained from the recurrence relation of the Chebyshev polynomial in the second equality. Using the recurrence relation of the Chebyshev polynomial, $U_{-1}(x)=0$ and the eigenvalue condition, we finally obtain  
\begin{align*}
||\mathbf{v}_{\lambda }||^{2}
&=2\left\{\frac{(1-\lambda ^{2})S_{n}}{q}+\frac{\lambda ^{2}-p}{q}\tilde{U}_{n-1}^{2}(\lambda /\sqrt{pq})-\frac{p}{q}\tilde{U}_{n-2}^{2}(\lambda /\sqrt{pq})\right\}\\
&=2\left\{\frac{(1-\lambda ^{2})S_{n}}{q}+\frac{\lambda ^{2}-(p+q)}{q}\tilde{U}_{n-1}^{2}(\lambda /\sqrt{pq})\right\}\\
&=\frac{2(1-\lambda ^{2})S_{n-1}}{q}.
\end{align*}

The eigen space of the time evolution operator $U^{(2n)}$ is described by that of $J_{2n}(p)$ (Lemma 2 of \cite{IdeKonnoSegawa2012}) as follows:
\begin{lem}\label{lem:eigenU}
Let $\lambda _{k}\ (k=1,\ldots, 2n)$ be the eigenvalues of $J_{2n}(p)$ and $\mathbf{v}_{\lambda _{k}}\ (k=1,\ldots, 2n)$ be the corresponding eigenvectors. We set $\lambda _{1}=1$ and $\lambda _{2n}=-1$. Then the eigenvalues $\mu _{k}\ (k=1, \pm 2,\ldots, \pm (2n-1), 2n)$ and corresponding eigenvectors $\mathbf{u} _{k}\ (k=1, \pm 2,\ldots, \pm (2n-1), 2n)$ of $U^{(2n)}$ are obtained as follows:

Let $\tilde{\mathbf{v}}_{\lambda }=\mathbf{v}_{\lambda }/||\mathbf{v}_{\lambda }||$ and for $k=1,\ldots 2n$, 
\begin{align*}
\mathbf{a}_{\lambda _{k}}&=
\tilde{\mathbf{v}}_{\lambda _{k}}(1)\ket{1,R}\\
&+\sum_{i=2}^{n}\tilde{\mathbf{v}}_{\lambda _{k}}(i)\ket{i}\otimes (\sqrt{p}\ket{L}+\sqrt{q}\ket{R})
+\sum_{i=n+1}^{2n-1}\tilde{\mathbf{v}}_{\lambda _{k}}(i)\ket{i}\otimes (\sqrt{q}\ket{L}+\sqrt{p}\ket{R})\\
&+\tilde{\mathbf{v}}_{\lambda _{k}}(2n)\ket{2n,L},\\
\mathbf{b}_{\lambda _{k}}&=S^{(2n)}\mathbf{a}_{\lambda _{k}}.
\end{align*}
Then we have

\begin{tabular}{l l}
$\mu _{1}=1$, & $\mathbf{u}_{1}=\mathbf{a}_{\lambda _{1}}=\mathbf{a}_{1}$,\\
$\mu _{\pm k}=\exp(\pm i\varphi_{\lambda _{k}})$, & 
$\mathbf{u}_{\pm k}=\mathbf{a}_{\lambda _{k}}-\exp(\pm i\varphi_{\lambda _{k}})\mathbf{b}_{\lambda _{k}}$, where $\cos\varphi_{\lambda _{k}}=\lambda _{k}$, for $k=2,\ldots ,2n-1$,\\
$\mu _{2n}=-1$, & $\mathbf{u}_{2n}=\mathbf{a}_{\lambda _{2n}}=\mathbf{a}_{-1}$.
\end{tabular}
\end{lem}

Now we estimate the distribution $\overline{p}_{i}^{(2n)}$. By the assumption of the choice of the initial state, we have 
\begin{eqnarray*}
\overline{p}_{i}^{(2n)}(x)=
\begin{cases}
\displaystyle\lim _{T\to \infty }\frac{1}{T}\sum_{t=0}^{T-1}
\left\lVert \left(\ket{x}\bra{x}\otimes I_{2}\right)\left(U^{(2n)}\right)^{t}(\ket{0}\otimes \ket{R})\right\rVert^{2},
&\text{if $i=1$,}\\
\displaystyle\lim _{T\to \infty }\frac{1}{T}\sum_{t=0}^{T-1}
\left\{
\frac{1}{2}
\sum_{J=L,R}
\left\lVert \left(\ket{x}\bra{x}\otimes I_{2}\right)\left(U^{(2n)}\right)^{t}(\ket{i}\otimes \ket{J})\right\rVert^{2}
\right\},
&\text{if $2\leq i\leq 2n-1$,}\\
\displaystyle\lim _{T\to \infty }\frac{1}{T}\sum_{t=0}^{T-1}
\left\lVert \left(\ket{x}\bra{x}\otimes I_{2}\right)\left(U^{(2n)}\right)^{t}(\ket{n+1}\otimes \ket{L})\right\rVert^{2},
&\text{if $i=2n$.}
\end{cases}
\end{eqnarray*}
Using the spectral decomposition $\left(U^{(2n)}\right)^{t}=\sum _{k}\mu _{k}^{t}\mathbf{u}_{k}\mathbf{u}_{k}^{\dag}$ and $\lim _{T\to \infty }(1/T)\sum_{t=0}^{T-1}e^{i\theta t}=\delta _{0}(\theta )\ (\text{mod} \ 2\pi)$, we obtain 
\begin{eqnarray*}
\overline{p}_{i}^{(2n)}(x)=
\begin{cases}
\displaystyle\sum _{k}\left\{(|u_{x,L}^{(k)}|^{2}+|u_{x,R}^{(k)}|^{2})\times |u_{1,R}^{(k)}|^{2}\right\},
&\text{if $i=1$,}\\
\displaystyle\frac{1}{2}
\sum _{k}\left\{(|u_{x,L}^{(k)}|^{2}+|u_{x,R}^{(k)}|^{2})\times (|u_{i,L}^{(k)}|^{2}+|u_{i,R}^{(k)}|^{2})\right\},
&\text{if $2\leq i\leq 2n-1$,}\\
\displaystyle\sum _{k}\left\{(|u_{x,L}^{(k)}|^{2}+|u_{x,R}^{(k)}|^{2})\times |u_{2n,L}^{(k)}|^{2}\right\},
&\text{if $i=2n$,}
\end{cases}
\end{eqnarray*}
with $\mathbf{u}_{k}=\sum _{j=1}^{2n}\ket{j}\otimes \left(u_{x,L}^{(k)}\ket{L}+u_{x,R}^{(k)}\ket{R}\right)$, because all eigenvalues of $U^{(2n)}$ are nondegenerate (This comes from nondegenerateness of $J_{2n}(p)$, see e.g.\ \cite{HoraObata2007}). Using this fact and Lemmas \ref{lem:eigenJacobi} and \ref{lem:eigenU}, we have the symmetric property of the time averaged distribution. Moreover we also obtain
\begin{align*}
|u_{i,L}^{(1)}|^{2}+|u_{i,R}^{(1)}|^{2}
=
|u_{i,L}^{(2n)}|^{2}+|u_{i,R}^{(2n)}|^{2}
=
|\tilde{\mathbf{v}}_{\pm 1}(i)|^{2}
=
\frac{ (p-q)[\delta _{1}(i)+\{1-\delta _{1}(i)\}/q](q/p)^{i-1} }{ 2\{2p-(q/p)^{n-1}\} },
\end{align*}
for $i=1,\ldots n$. On the other hand, it is obvious that
\begin{eqnarray*}
\overline{p}_{i}^{(2n)}(x)\geq 
\begin{cases}
\displaystyle\sum _{k=1,2n}\left\{(|u_{x,L}^{(k)}|^{2}+|u_{x,R}^{(k)}|^{2})\times |u_{1,R}^{(k)}|^{2}\right\},
&\text{if $i=1$,}\\
\displaystyle\frac{1}{2}
\sum _{k=1,2n}\left\{(|u_{x,L}^{(k)}|^{2}+|u_{x,R}^{(k)}|^{2})\times (|u_{i,L}^{(k)}|^{2}+|u_{i,R}^{(k)}|^{2})\right\},
&\text{if $2\leq i\leq n$.}
\end{cases}
\end{eqnarray*}
As a consequence, we have the desired lower bound. 
\begin{align*}
\bar{p}_{i}^{(2n)}(x)\geq
\frac{(p-q)^{2}(q/p)^{i-1}}{2\left[\delta _{1}(i)+\left\{1-\delta _{1}(i)\right\}2q\right]\left\{2p-(q/p)^{n-1}\right\}^{2}}\left[\delta _{1}(x)+\{1-\delta _{1}(x)\}\frac{(q/p)^{x-1}}{q}\right].
\end{align*}
Taking suitable limits, the remaining parts of Theorem \ref{thmtimeavedist} are obtained by this lower bound.

The eigenvectors $\mathbf{u}_{\pm k}$ with $k=2,\ldots ,2n-1$ are not obtained explicitly in the present stage. It is expected that the remaining part of the time averaged distribution converges to the uniform distribution after suitable normalization because the remaining eigenvectors are similar to that of the time evolution operator of DTQW on the path graphs \cite{IdeKonnoSegawa2012}. It is an  interesting future work to make it clear whether this conjecture is true or not. 
%%%%%%%%%%%%%%%%%%%%%%%%%%%%%%%%%%%%%%%%%%
%
%\section{Conclusions}
%
%
%%%%%%%%%%%%%%%%%%%%%%%%%%%%%%%%%%%%%%%%%%
\par
\
\par\noindent
{\bf Acknowledgments.} 
Y. I. was supported by the Grant-in-Aid for Young Scientists (B) of Japan Society for the Promotion of Science (Grant No.\ 23740093). N. K. was supported by the Grant-in-Aid for Scientific Research (C) of Japan Society for the Promotion of Science (Grant No.\ 24540116). E.S. was supported by the Grant-in-Aid for Young Scientists (B) of Japan Society for the Promotion of Science (Grant No.\ 25800088). X.-P. X. was supported by the National Natural Science Foundation of China under project 11205110.

\begin{small}

\end{small}

\end{document}